\documentclass[%
reprint,
superscriptaddress,
 amsmath,amssymb,
 aps,
prb,
]{revtex4-2}

\usepackage{dcolumn}% Align table columns on decimal point
\usepackage{bm}% bold math
\usepackage{amsmath}
\usepackage{amssymb}
\usepackage{graphicx}
\usepackage{bm}
\usepackage{gensymb}
\usepackage{braket}
\usepackage{xcolor,soul}
\usepackage{float}
\usepackage[colorlinks=true, allcolors=blue]{hyperref}

\newcommand{\bk}[0]{\mathbf{k}}
\newcommand{\smallQi}{\scriptscriptstyle \mathbf{Q}_i}
\begin{document}

\preprint{APS/123-QED}

\title{Chiral Magnetism and Quantum Anomalous Hall Effect in a Low-energy Kondo Model on the Triangular Lattice}

\author{Kai Vylet}
\affiliation{Department of Physics, University of California Santa Barbara, CA 93106, USA}
\author{Xingkai Huang}
\affiliation{Department of Physics, University of California Santa Barbara, CA 93106, USA}
\author{Leon Balents}
\affiliation{Kavli Institute for Theoretical Physics, University of California, Santa Barbara, CA 93106, USA}
\affiliation{French American Center for Theoretical Science, CNRS, KITP, Santa Barbara, California 93106-4030, USA}
\affiliation{Canadian Institute for Advanced Research, Toronto, Ontario, Canada}

\date{\today}

\begin{abstract}
We study an effective low-energy Kondo model on the triangular lattice in which itinerant electrons occupy a valence pocket at $\Gamma$ and three conduction pockets at the $M$ points of the Brillouin zone. This construction has a Fermi-surface nesting structure that favors triple-$Q$ magnetic order while only assuming the low-energy band-structure. Treating the local moments as classical spins on a four-sublattice magnetic unit cell, we find extended regions of non-coplanar order, including tetrahedral and related canted tetrahedral states, in addition to ferromagnetic and coplanar phases. The chiral phases remain stable over a broad range of inter-pocket Kondo couplings and persist in the presence of an external magnetic field. For certain chiral orders, the electronic bands can become gapped and host a quantum anomalous Hall state with $\sigma_{xy}=4\,e^2/h$. These results show that chiral magnetism and a quantized anomalous Hall effect on the triangular lattice do not rely on a specific tight-binding band structure, but can arise more generally from low-energy nested pockets at $\Gamma$ and $M$.
\end{abstract}

\maketitle

\section{Introduction}

The recent experimental realization of intrinsic magnetism in atomically thin crystals has greatly expanded the potential toolkit of two-dimensional (2D) materials beyond their already numerous uses~\cite{Zhang2024,Gibertini2019,Jiang2021}. Besides the difference in dimensionality with bulk materials, these 2D materials allow for the precise engineering of magnetic phenomena through layer-dependent properties and external stimuli, such as strain and electric gating.
Magnetic quasi-2D materials and heterostructures that host topological phenomena, such as the quantum anomalous Hall (QAH) effect, are also highly desirable for realizing dissipationless electron transport and fault-tolerant topological quantum computing~\cite{Deng2020}. Many of these uses take advantage of complex, non-collinear spin textures. 

An example of a quasi-2D system displaying such physics is the van der Waals material GdGaI, which has recently been found to have non-coplanar/chiral magnetic orders, e.g., tetrahedral, and large anomalous Hall conductivity at low temperature~\cite{okuma2024emergenttopologicalmagnetismhunds}. In this material, a triangular lattice of Gd atoms provides fixed, classically-natured spin-7/2 moments while the semiconductor layers provide mobile electrons. A DFT calculation of the electronic band structure shows electrons concentrated at the $M$ points of the Brillouin zone (BZ) and holes around $\Gamma$.

The presence of these two ingredients, a lattice of classical spins and itinerant electrons, naturally points to known explanations of chiral spin order and electronic topology via the Kondo or Hund's interaction~\cite{hayamiEffectiveBilinearbiquadraticModel2017,Ozawa2016, Chern2010}. In close relation, previous work on the triangular lattice with a 4-site magnetic unit cell has shown that a tetrahedral spin configuration is stabilized via Kondo interaction with tight-binding electrons at 3/4 filling~\cite{Martin2008,Kato2010,Akagi2010}. Specifically, a weak-coupling instability is generated by Fermi surface nesting via the three $\vec M_i$ momentum vectors in the first Brillouin zone (BZ). The electrons can furthermore form a quantum anomalous Hall (QAH) state with Hall conductivity $\sigma_{xy}=e^2/h$.

However, the example of GdGaI suggests that chiral spin order and anomalous Hall physics on the triangular lattice can arise more generally outside of a tight-binding approximation. In this work, we investigate an effective Kondo model on the triangular lattice which has its low-energy electronic states around the $M$ and $\Gamma$ points, like GdGaI. Similar to the 3/4-filled tight-binding model, such a setup has Fermi surface nesting by the $\vec M_i$ vectors, but makes no assumptions on the higher energy states or filling factor. We find that the model supports chiral and non-chiral spin phases, both in zero and finite field. Furthermore, tetrahedral and `tetrahedral-like' configurations can generate a QAH state with $\sigma_{xy}=4\,e^2/h$, which can be understood as coming from two band inversions each contributing Chern number $C=2$ (see App.~\ref{app: Chern number} for details).

\section{Model}

\begin{figure}[htbp]{}
    \centering
    \includegraphics[scale=0.42]{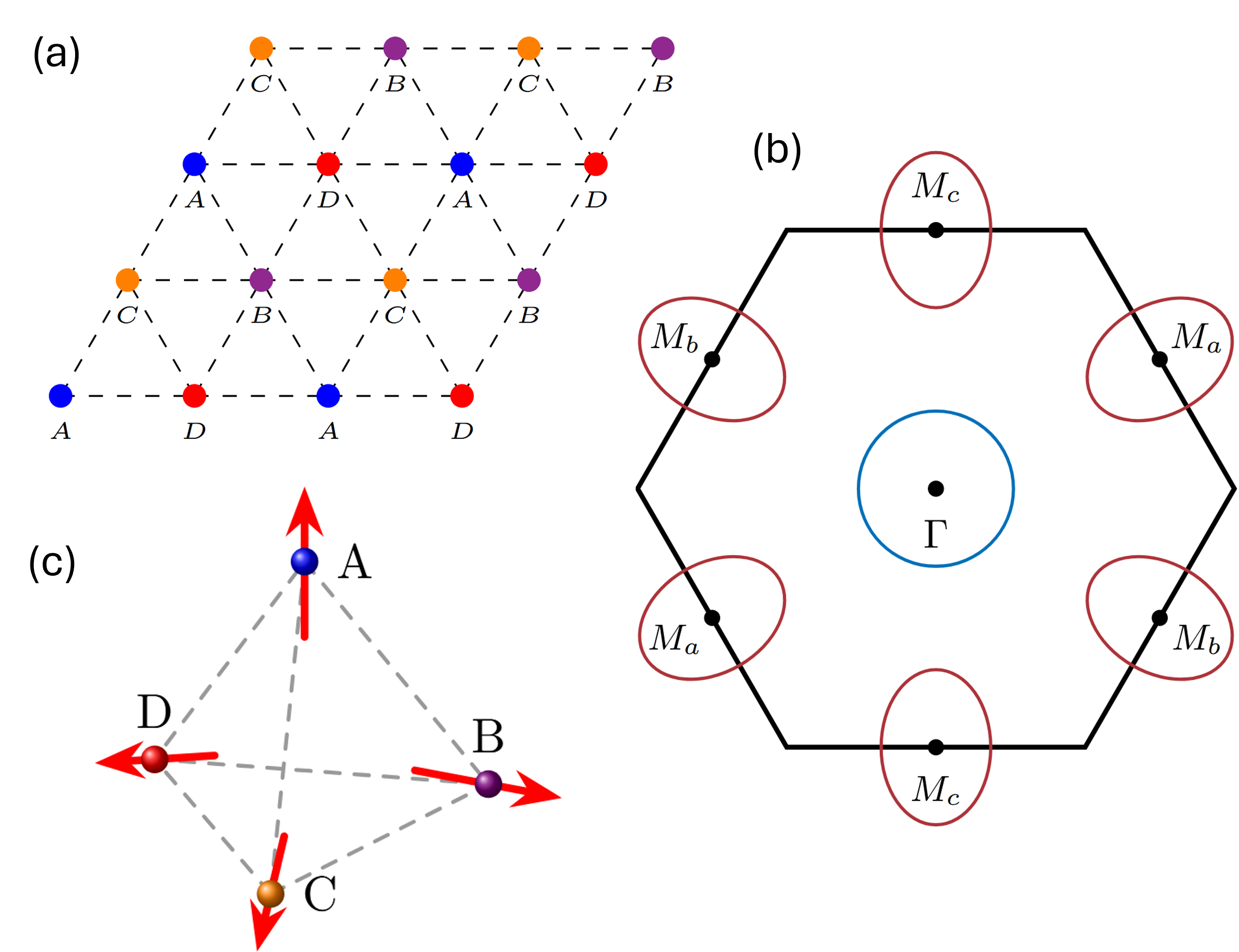}
    \caption{(a) Triangular lattice with a 4-site magnetic unit cell with sites $\{A,B,C,D\}$. We set the lattice constant to unity in subsequent calculations and plots. (b) First Brillouin zone of the triangular lattice. The low-energy valence and conduction states in the model lie around $\Gamma$ and the $M$ points $\{M_a,M_b,M_c\}$, respectively; the circle and ellipses depict cross sections of the itinerant electron dispersions, given in Eq.~\eqref{eq: dispersions}. (c) An example of an all-out tetrahedral spin configuration, in which each spin points outwards from the center of a tetrahedron. The configuration forms a chiral spin state with zero net magnetization.}
    \label{fig: lattice and BZ}
\end{figure}

The model takes place on a triangular lattice with a 4-site magnetic unit cell, depicted in Fig.~\hyperref[fig: lattice and BZ]{\ref*{fig: lattice and BZ}a}. We implement the GdGaI scenario by setting the low-energy states to live in localized ``pockets" near the $M$ and $\Gamma$ points in the BZ, as illustrated in Fig.~\hyperref[fig: lattice and BZ]{\ref*{fig: lattice and BZ}b}; here, we choose the $M$ points, labeled $\{M_a,M_b,M_c\}$, to host conduction bands and the $\Gamma$ point to host a valence band. For generality, the conduction bands are given some tuneable ellipticity. 
 
The Kondo interaction on a lattice model is typically written as $H_K=J\sum_r\vec{S}_r\cdot\frac{1}{2}\psi_r^\dagger\vec\sigma\psi_r$, where $\psi^\dagger_r$ is the electron creation operator at site $r$, $\vec S_r$ is the spin fixed at the site, and $J$ is the Kondo coupling. We implement our low-energy description of the electronic states by writing the electron operators in momentum space and truncating the higher energy modes:
\begin{align}
    \psi_r\sim \frac{1}{\sqrt{N}}\sum_{\mathbf{Q}_i}\sum_{|\bk|\leq\Lambda}e^{i(\smallQi+\bk)\cdot\vec{r}}\,c_{\smallQi+\bk}\,,
\end{align}
where the $\mathbf{Q}_i$ denote the pocket locations $M_i$ and $\Gamma$, and $c^\dagger,\,c$ are the Bloch electron ladder operators. This ansatz picks out the modes localized around the pockets by setting a cutoff scale on the deviations from the pockets: $|\bk|\leq\Lambda$, which we assume is larger than other energy scales in the model. Inputting this ansatz into $H_K$ and using the four-sublattice structure, $H_K$ takes the form

\begin{align}
\label{eq: Kondo term}
    {H}_{K}=\sum_I \sum_{\mathbf k,\sigma,\sigma'}\sum_{i,j=v,a,b,c} \vec S_I \cdot J^{ij}_I\, c_{i,\mathbf k\sigma}^\dagger\, \vec\sigma_{\sigma\sigma'}\,c_{j,\mathbf k \sigma'}\,,
\end{align}
where we have absorbed extra sublattice and momentum-dependent factors into new couplings $J_I^{ij}$. We have relabeled the ladder operators to $c_{i,\mathbf k\sigma}$ so that the pocket type $\mathbf{Q}_i$ is now denoted by an index (valence by $i=v$ and conduction type by $i=a,b,c$); the electron spin indices $\sigma,\sigma'$ are now also explicit. Again, one should think of the momentum $\bk$ as a small deviation away from the corresponding $M_i$ or $\Gamma$ point in the Brillouin zone. Lastly, $\vec S_I$ is the spin on sublattice $I\in\{A,B,C,D\}$, which we normalize to $|\vec S_I|^2=1$.  In this effective description, the Kondo terms mediate intra- and inter-pocket scattering.

Because we only describe the low-energy physics, we treat the $J_I^{ij}$ as phenomenological, independent couplings and will refer to them as the Kondo couplings of the model. One can then impose lattice symmetries in the Hamiltonian to reduce to four independent couplings: packaging $J_I^{ij}$ into a matrix form, we have the $I=A$ couplings to be
\begin{align}
J_A=
    \begin{pmatrix}
    J_{v} & J_{vc} & J_{vc} & J_{vc}\\  
    J_{vc} & J_{c} & J_{cc} & J_{cc} \\ 
    J_{vc} & J_{cc} & J_{c} & J_{cc} \\ 
    J_{vc} & J_{cc} & J_{cc} & J_{c} \\ 
    \end{pmatrix}\,,
\end{align}
where we have renamed the couplings to $\{J_v\,,J_{c}\,,J_{vc}\,,J_{cc}\}$ to emphasize that there are four independent ones. Here, the column indices correspond to the pocket indices in the order $\{v,a,b,c\}$. Valence to valence scattering is mediated by $J_v$ ($J_A^{vv}$ before), a conduction pocket to itself by $J_c$ (e.g., $J_A^{aa}$ before), valence to conduction by $J_{vc}$ (e.g., $J_A^{va}$ before), and between different conduction pockets by $J_{cc}$ (e.g., $J_A^{ab}$ before). Note that the latter three couplings are now independent of conduction pocket types. The other coupling matrices $J_B$, $J_C$, and $J_D$ have the same components as $J_A$ up to minus signs. Calculation of the coupling matrices is detailed in Appendix~\ref{app: matrix constraints}.

With the Kondo interactions in hand, we now construct the Hamiltonian by adding in the itinerant electron dispersions: 
\begin{align}
\label{eq: Hamiltonian}
H = H_K+\sum_{\mathbf{k},\sigma} \sum_{i=v,a,b,c}  \varepsilon_{i}(\mathbf k)\,c_{i, \mathbf{k}\sigma}^\dagger c_{i, \mathbf{k}\sigma}\,,
\end{align}
where we use the form of $H_K$ in Eq.~\eqref{eq: Kondo term}. The valence dispersion $\varepsilon_v(\mathbf k)$ and conduction dispersions $\varepsilon_i(\mathbf k)$ are given by
\begin{align}
    \label{eq: dispersions}
    &\varepsilon_v(\mathbf k)=\epsilon_v-\frac{1}{2m_h}\big(k_x^2+k_y^2\big)\,,\\\nonumber
    &\varepsilon_i(\mathbf k) =\epsilon_c + \frac{1}{2m_e} \bigg[(R_{\theta_i} \mathbf{k})_x^2 + \alpha (R_{\theta_i} \mathbf{k})_y^2 \bigg]\,.
\end{align}
The parameters $\epsilon_c$ and $\epsilon_v$ set the bottom and top of the conduction and valence bands, respectively, while $m_e$ and $m_h$ set the effective masses of the dispersions. $R_{\theta_i}$ is the 2D rotation matrix with angle $\theta_i$, and $\alpha$ is a number that determines the ellipticity of the conduction dispersions ($\alpha=1$ gives an isotropic dispersion). To maintain the lattice rotation symmetry, we rotate the conduction dispersions $60\degree$ relative to each other by setting $\theta_a = \frac{\pi}{6} ,  \theta_b = -\frac{\pi}{6}$, and $\theta_c = \frac{\pi}{2}$.

Due to the lattice translation symmetry, the model is solvable using Bloch's theorem. We can reformat Eq.~\eqref{eq: Hamiltonian} to make the Bloch Hamiltonian apparent:
\begin{align}
&H = \\\nonumber
&\sum_{\mathbf k} 
\begin{pmatrix}
\Phi_{\mathbf k,\uparrow}^\dagger\,\,\Phi_{\mathbf k,\downarrow}^\dagger
\end{pmatrix}
\left(\sigma_0\otimes\mathcal{T}({\mathbf k}) + \sum_{I=A}^D \vec S_I\cdot \vec \sigma \otimes J_I\right)
\begin{pmatrix}
\Phi_{\mathbf k,\uparrow}\\\Phi_{\mathbf k,\downarrow}
\end{pmatrix}
\end{align}
where $\Phi_{\mathbf k,\sigma}\equiv(c_{v,\mathbf k\sigma}\,, c_{a,\mathbf k\sigma}\,, c_{b,\mathbf k\sigma}\,, c_{c,\mathbf k\sigma}\,)$, $\mathcal{T}(\mathbf k)\equiv\text{Diag}[\varepsilon_v(\mathbf k)\,,\varepsilon_a(\mathbf k)\,,\varepsilon_b(\mathbf k)\,,\varepsilon_c(\mathbf k)]$, and the $8\times8$ Bloch Hamiltonian is $\mathcal{H}(\mathbf k)\equiv \sigma_0\otimes\mathcal{T}({\mathbf k}) + \sum_{I=A}^D \vec S_I\cdot \vec \sigma \otimes J_I$.

In the case of zero average magnetization, $\vec S_\text{avg}=1/4\sum_{I=A}^D \vec S_I=0$, there is a two-fold degeneracy of the bands. To see this, we first note that a global rotation of the classical spins amounts to a similarity transform of the Hamiltonian. Using this freedom, we can parametrize any $\vec S_\text{avg}=0$ configurations as $\vec S_A=(S^x,S^y,S^z),\,\vec S_B=(-S^x,-S^y,S^z),\,\vec S_C=(-S^x,S^y,-S^z),$ and $\vec S_D=(S^x,-S^y,-S^z)$, with $|\vec S_A|^2=1$ (see App.~\ref{app: spin parametrization} for details). Upon substituting this parametrization, one finds that the Bloch Hamiltonian can be split into two disjoint blocks:
\begin{align}
    \label{eq: Hamiltonian blocks}
    \mathcal{H} = c_I^\dagger\,\hat{\mathcal{H}}_0\,c_I + c_{II}^\dagger\,\hat{\mathcal{H}}_0\,c_{II}\,,
\end{align}
where $c_I\equiv(c_{v,k\uparrow}\,,c_{a,k\uparrow}\,,c_{b,k\downarrow}\,,c_{c,k\downarrow})$, and $c_{II}\equiv(c_{v,k\downarrow}\,,-c_{a,k\downarrow}\,,-c_{b,k\uparrow}\,,c_{c,k\uparrow})$, with 
\begin{align}
\hat{\mathcal{H}}_0\equiv
    \begin{pmatrix}
    \varepsilon_v(\bk) & 4 J_{vc}\,S^{z} & -4 iJ_{vc}\,S^{y} &  4 J_{vc}\,S^{x}\\
    4 J_{vc}\,S^{z} & \varepsilon_a(\bk) & 4 J_{cc}\,S^{x} & -4 i J_{cc}\,S^{y} \\
    4 iJ_{vc}\,S^{y} &  4 J_{cc}\,S^{x} & \varepsilon_b(\bk) & -4 J_{cc}\,S^{z}\\
    4 J_{vc}\,S^{x} & 4 i J_{cc}\,S^{y} & -4 J_{cc}\,S^{z} & \varepsilon_c(\bk)
\end{pmatrix}\,.
\label{eq: H up}
\end{align}
Notably, the Hamiltonian's dependence on $J_{v}$ and $J_c$ drops out. Physically, this double degeneracy is due to an additional symmetry of $\mathcal H$~\cite{Martin2008}. First, a translation by lattice vector $\vec a_1=(1,0)$ exchanges $\vec S_A\leftrightarrow\vec S_D$, $\vec S_B\leftrightarrow\vec S_C$ and sends $(c_{a,k},\,c_{b,k})\to(-c_{a,k},\,-c_{b,k})$. Next, performing a $\pi$ spin rotation about the $x$-axis restores the classical spin state while sending $(c_{i,\uparrow},\,c_{i,\downarrow})\to(-ic_{i,\downarrow},\,-ic_{i,\uparrow})$. Composing this translation and spin rotation then leaves $\hat{ \mathcal{H}}_0$ invariant while exchanging the two sectors up to an overall phase, $c_I\leftrightarrow -i\, c_{II}$.

As we will see in the spin ground state calculations, the tetrahedral configuration, illustrated in Fig.~\hyperref[fig: lattice and BZ]{\ref*{fig: lattice and BZ}c}, turns out to be a prevalent non-coplanar, zero-magnetization spin state in this model.

\section{Spin Ground States}
\label{sec: spin ground states}

\begin{figure*}[htbp]
    \centering
    \includegraphics[width=1\linewidth]{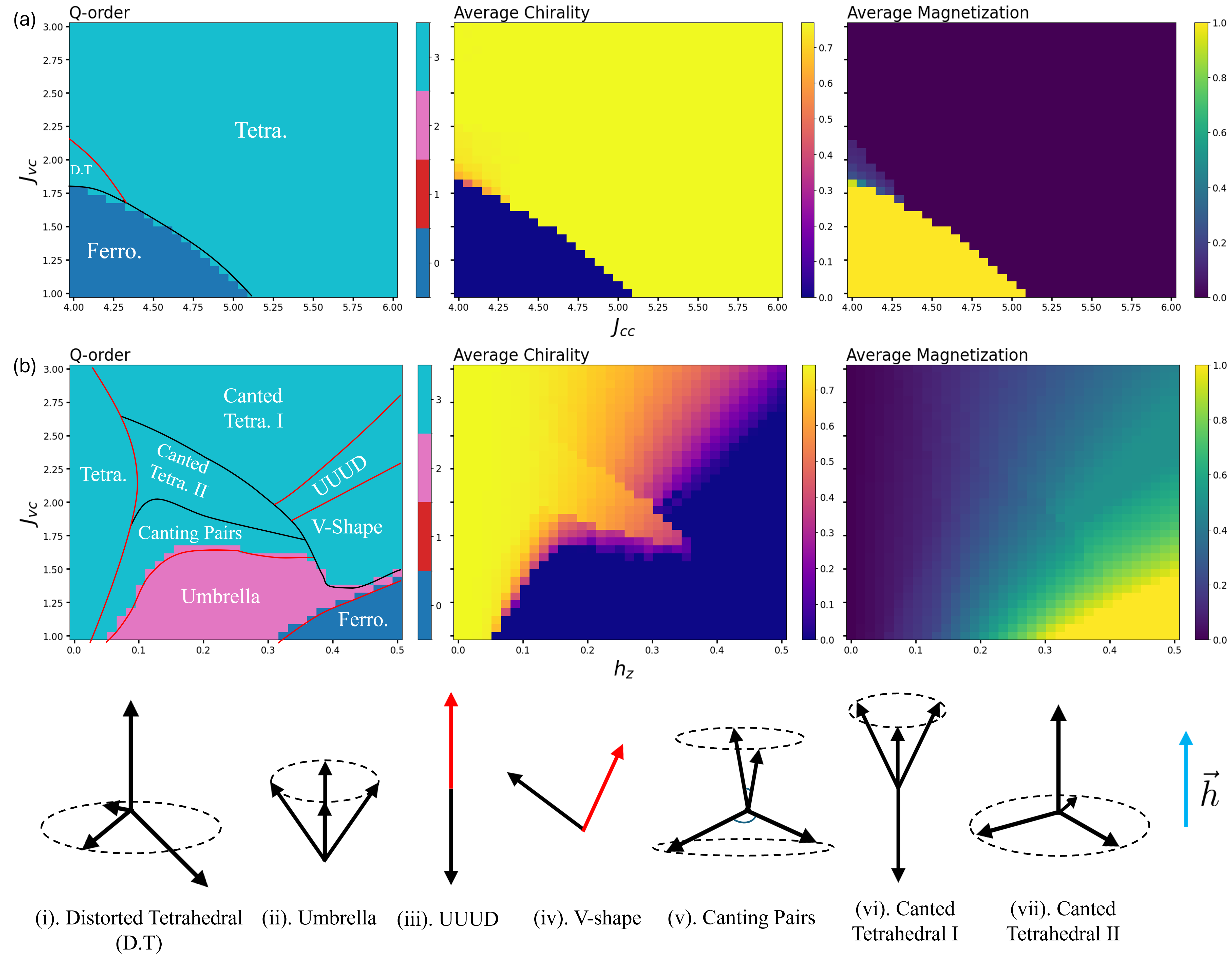}
    \caption{The $q$-ordering, average scalar chirality, and average magnetization of the ground states as $J_{vc}$ and, (a) $J_{cc}$ or (b) an external field $\vec h=h_z\,\hat z$, are varied. The other couplings are held fixed to $(J_v, J_c)=(2,4)$ in both cases and $J_{cc}=6$ in (b). Black and red lines denote first and second order phase transitions, respectively. In both diagrams, the states with largest average chirality, zero magnetization, and $3q$ characterization are tetrahedral. Other chiral and non-chiral $3q$ states with net magnetization appear as well. Illustrations of the phases are shown in (i)-(vii); the red arrows identify when three of the spins are degenerate. Here, the Distorted Tetrahedral state is given by lowering (raising) one leg of the tetrahedral configuration while raising (lowering) the other two to a lesser degree, while the Canted Tetrahedral I and II have all three legs cant upwards equally towards the field. In the Canting Pairs state, the upwards and downwards pairs cant at different angles and are perpendicular, e.g., one can think of the upper pair as projecting onto the $x$-axis and the lower pair onto the $y$-axis.}
    \label{fig: Spin phase diagram 1}
\end{figure*}

We first study the spin ground states appearing in the model as a function of the Kondo couplings, assuming a 4-site magnetic unit cell. To begin, one may heuristically expect that regimes where the inter-pocket couplings $(J_{vc}, J_{cc})$ are much larger than the intra-pocket couplings $(J_v,J_c)$ are likely to support $3q$ states, due to Fermi surface nesting. On the other hand, ferromagnetic behavior would be expected in regimes with larger $(J_v,J_c)$, since zero wave-vector scattering would dominate.

To determine the spin ground state, we fix the number of electrons in the system, vary the spin state, and calculate the total occupied electron energy from the band structure; the spin ground state is then identified by whichever spin state minimizes this energy. Notably, the spin state space is continuous due to the classical nature of the spins. We search this space for a global energy minimum using a differential evolution algorithm, which is a population-based stochastic method that iteratively improves candidate solutions~\cite{Storn1997}. In each iteration, new candidate solutions are generated by combining existing ones using mutation and crossover, and they are retained if they have a lower energy. In our implementation, the differential evolution algorithm was set with a population size of $N_p = 20$, a maximum number of generations $N_{\text{gen}} = 300$, and a convergence tolerance of $10^{-3}$. To get the final phase diagram, we then perform additional refinements at each point by comparing the differential evolution result against a local minimization search around neighboring ground states as well as several ansatze. Unless otherwise noted, we set the dispersion parameters to $\epsilon_c=-5,\,\epsilon_v=5,\,m_e=m_h=0.015,$ and $\alpha=2$. 

We characterize the spin ground states by their $q$-ordering, average scalar spin chirality, and average magnetization. The scalar spin chirality on a plaquette is given by
\begin{align}
\chi_{ijk}\equiv\vec S_i\cdot(\vec S_j\times\vec S_k)\,,
\end{align}
and to capture the chirality of our 4-site spin state, we report the magnitude of the scalar chirality averaged over the four types of plaquettes: $|\langle \chi \rangle| \equiv \frac{1}{4}| \chi_{ADC} + \chi_{BCD} + \chi_{ABD} + \chi_{ACB} \bigr|$. By $q$-ordering, we mean that the spin pattern can be decomposed into Fourier modes
\begin{align}
\label{eq: q decomposition}
  \vec{S}_r=\vec{A}_0\,+\sum_{j=a,b,c} \vec{A}_j\, e^{i\vec{M}_j\cdot \vec{r} }\,,
\end{align}
where $\vec M_j$ are the $M$ point momentum vectors of the BZ. A state is said to be triple-$Q$, or $3q$, if three of the $\vec A_j$ are non-zero. 2$q$ and $1q$ states are defined similarly, and we denote a ferromagnetic state, $\vec S_r=\vec A_0$, as $0q$.

Fig.~\hyperref[fig: Spin phase diagram 1]{\ref*{fig: Spin phase diagram 1}a} shows the spin phase diagram as a function of $J_{vc}$ and $J_{cc}$. The Fermi energy used in these calculations was set by requiring charge neutrality in the band structure when the Kondo couplings are set to zero, though we observe that the variety of states, including chiral states, are generic for various choices. In particular, non-coplanar states with significant scalar chirality turn out to be prevalent; the $3q$ states with the maximum scalar chirality and zero magnetization are tetrahedral. The tetrahedral regime transitions from the ferromagnetic one in two ways: a first-order transition or through a transition region of ``distorted-tetrahedral (D.T.)" states. 

A wider variety of phases appears when considering an external magnetic field in the system (taken to be in the $+\hat z$ direction), reducing the global spin rotation symmetry to just $SO(2)$ rotation about the $z$-axis. Such a phase diagram for fixed $J_{cc}=6$ is shown in Fig.~\hyperref[fig: Spin phase diagram 1]{\ref*{fig: Spin phase diagram 1}b}. Here, the tetrahedral and ferromagnetic regions are separated by other types of tetrahedral distortions as well as more familiar categories of classical spin states.

Details on the energy calculation and additional diagrams at $(J_v,J_c)=(0,0)$ are discussed in Appendix~\ref{app: external field}.

\begin{figure*}[!t]
        \centering
        \includegraphics[scale=0.42]{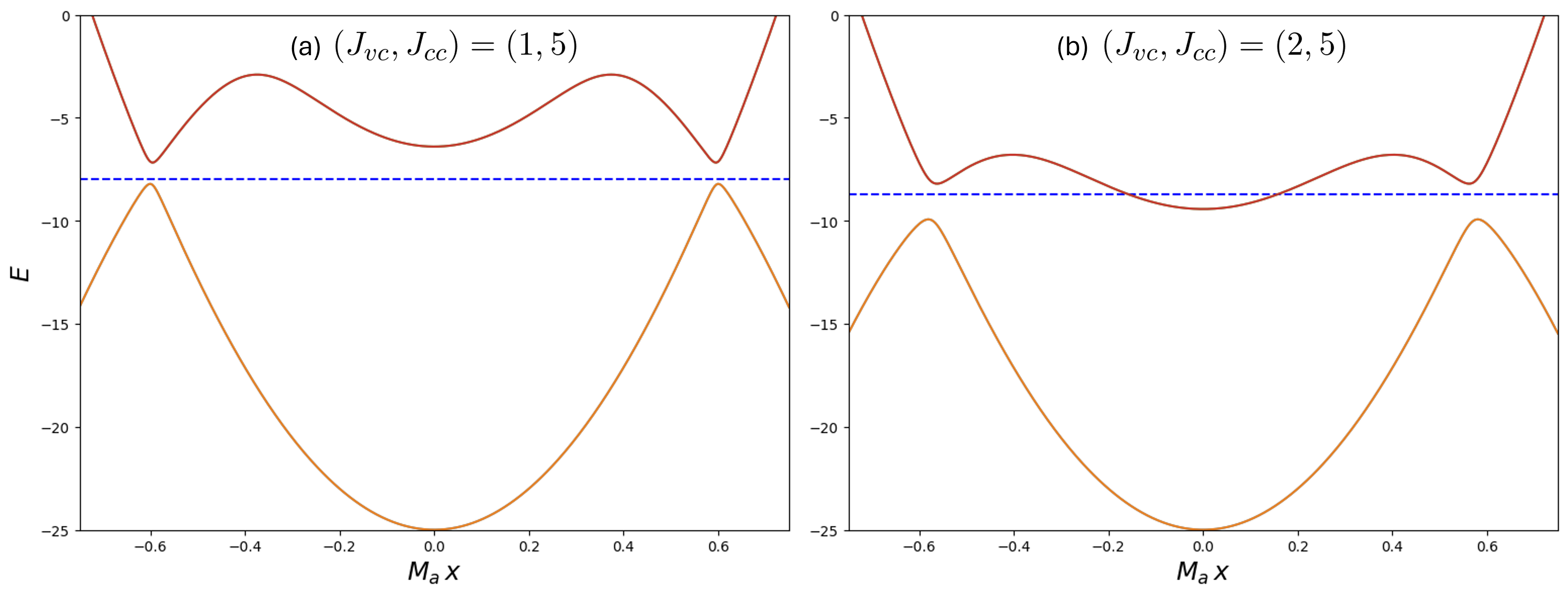}
        \caption{The lowest two sets of degenerate bands in $\bk$-space for the tetrahedral spin state. The horizontal axis is a line along the $\vec M_a\equiv (\pi,\pi/\sqrt{3})$ direction through $\bk=0$. For both parameter choices the band inversions occur, which opens a gap and generates a Chern number of $C=4$ in the lowest pair of degenerate bands. In (a), one can place the Fermi surface (dashed blue line) anywhere between the gap in the avoided crossings to achieve an insulating state. For the other parameter choice in (b), the upper bands lower into this gap. However, since the Berry curvature is concentrated around the inversion points (Fig.~\ref{fig: berry curvatures}), the metallic state with the illustrated Fermi surface appears to have a Hall conductivity of $\sigma_{xy}=4\, e^2/h$.}
    \label{fig: bands}
\end{figure*}

\section{Quantum Anomalous Hall Effect}

We find that electronic states generically display topological character in the presence of a chiral
spin configuration. Namely, when the net scalar spin chirality is non-zero, the system can display an anomalous
Hall effect. This is characterized by a non-zero Hall conductivity~\cite{Nagaosa2010}
\begin{align}
    \label{eq: Hall conductivity}
    \sigma_{xy}=-\frac{e^2}{\hbar}\sum_n\int_{|\bk|\leq\Lambda}\frac{d\mathbf k}{(2\pi)^2}\,n_F(\varepsilon_n(\bk))\,\Omega_n(\bk)\,,
\end{align}
with $\varepsilon_n(\bk)$ the $n^{\text{th}}$ band, $n_F$ the Fermi function at zero temperature and $\Omega_n(\bk)$ the Berry curvature of the $n^{\text{th}}$ band:
\begin{align}
    \Omega_n(\bk)=-2\,\text{Im}\,\sum_{m
    \neq n} \frac{\bra{n,\bk}\partial_{k_x}\mathcal{H}\ket{m,\bk}\bra{m,\bk}\partial_{k_y}\mathcal{H}\ket{n,\bk}}{[\varepsilon_n(\mathbf k)-\varepsilon_{m}(\bk)]^2}\,.
\end{align}
The momentum integral is taken within the cutoff region $|\bk|\leq\Lambda$, but we find generally that $\sigma_{xy}$ is cutoff-independent for large enough $\Lambda$. To avoid issues with band degeneracy in the zero-magnetization spin configurations, we calculate the Berry curvatures within each sector separately in such cases and sum the results.

With the presence of an AHE in the model, it is natural to ask whether chiral spin states with a gapped band structure can produce a quantized Hall conductivity. Indeed, this turns out to be possible. The clearest example is with a tetrahedral spin state, in which a gap opens between the lowest and second lowest set of degenerate bands for certain choices of the Kondo couplings, e.g., $(J_{vc},J_{cc})=(1,5)$ and $(2,5)$, as shown in Fig.~\ref{fig: bands}. Integrating the Berry curvature within the cutoff, one finds that the lowest band in each sector has Chern number $C=2$ and so together contribute Hall conductivity $\sigma_{xy}=4 \,e^2/h$; hence the system achieves an anomalous Quantum Hall state in its insulating phase. See App.~\ref{app: Chern number} for further discussions on the quantization value. We confirm that quantized conductivity in the model is not generally tied to the specific choices of $\{\epsilon_c,\, \epsilon_v\,, m_h\,, m_e\,, \alpha\}$, in that changes the parameters and gap size can be counteracted by adjusting the Kondo couplings $J_{vc}$ and $J_{cc}$ appropriately. An exception to this is the fine-tuned case of isotropic conduction dispersions, $\alpha=1$, where the Hall conductivity vanishes due to additional symmetry, as detailed in Appendix~\ref{app: isotropic}.

Fig.~\ref{fig: berry curvatures} shows that the Berry curvature for the lowest and second-lowest pairs of bands are concentrated near the avoided crossing, which spans over a circle in momentum space. This localization of the Berry curvature near $\bk=0$ allows $\sigma_{xy}$ to be insensitive to the momentum cutoff and hence supports the identification of a QAHE. Another interesting consequence of the Berry curvature distribution occurs in band structures similar to $(J_{vc},J_{cc})=(2,5)$. In such cases, the upper pair of bands dip into the energy gap between the inversion points, providing a metallic state where the lower bands are fully occupied and the upper bands are only occupied where the Berry curvature is negligible. Hence, this metallic state appears to have an approximately quantized Hall conductivity.  

\begin{figure*}[]
    \centering
    \includegraphics[scale=0.42]{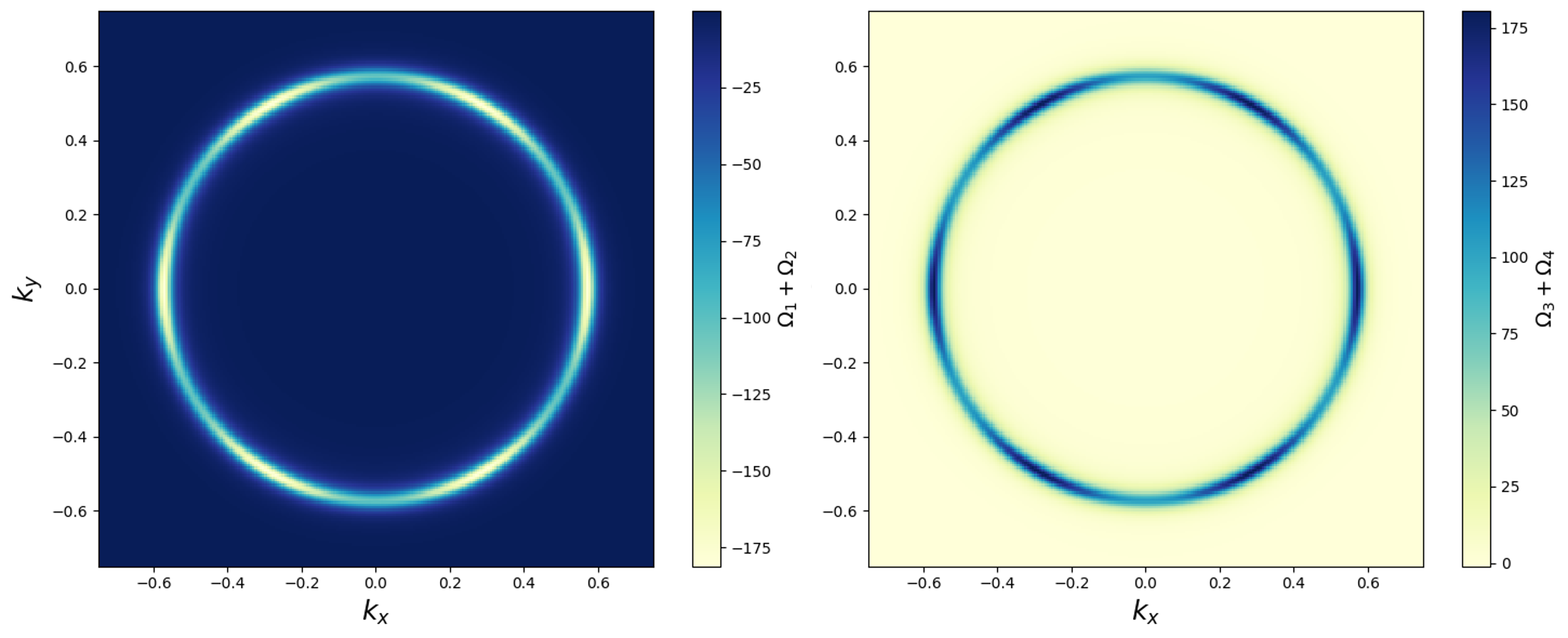}
    \caption{The total Berry curvature of the lowest two degenerate bands (left) and the next lowest degenerate bands (right), for the $(J_{vc},J_{cc})=(2,5)$ system shown in Fig.~\hyperref[fig: bands]{\ref*{fig: bands}b}. The curvature is concentrated around the avoided crossing, which occur along a ring of points in $\bk$-space.}
    \label{fig: berry curvatures}
\end{figure*}

\begin{figure*}[]
    \centering
    \includegraphics[width=\linewidth]{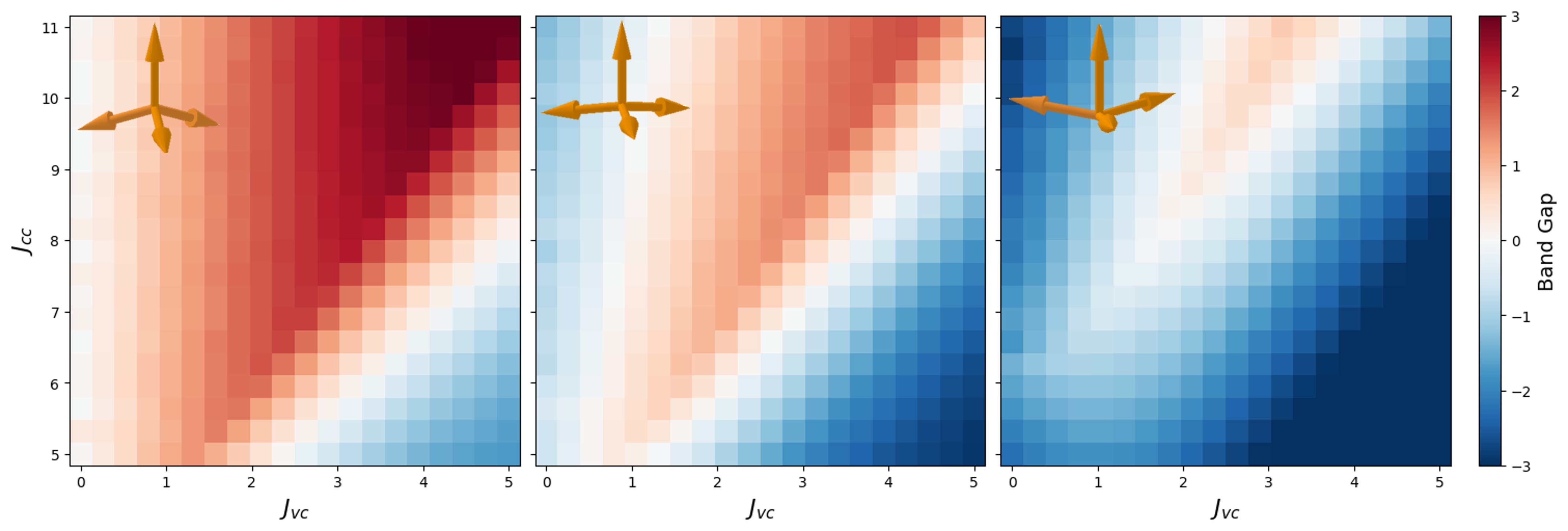}
    \caption{The energy gap between the lowest and second lowest sets of bands in the tetrahedral and canted tetrahedral band structures as $J_{cc}$ and $J_{vc}$ are varied. From left to right: the tetrahedral state, a deformation where the three downward-pointing spins are each canted $10\degree$ upwards in the $z$-direction, and a similar deformation but canted by $25\degree$.  A negative gap (blue) denotes a metallic state. Each of the spin states have non-zero chirality, and the points in the gapped regions host a $\sigma_{xy}=4\,e^2/h$ QAH state. The gap size at a given point decreases as the magnetization increases, and furthermore the gap survives only for points where $J_{vc}$ and $J_{cc}$ are large enough to counteract the diminishing magnitudes of the alternating spin sums.}
    \label{fig: gap diagram}
\end{figure*}

We note that the QAHE in the model is not restricted to the tetrahedral state, but could potentially appear for any chiral spin state that has a gapped band structure. However, we expect that the gap size is inversely related to the average magnetization of the classical spins. To see why, we note that the $J_{vc}$ and $J_{cc}$ terms in the Hamiltonian, i.e., the ones which mix different pockets, each depend on an alternating sum of the spin components: $S^\alpha_A+S^\alpha_B-S^\alpha_C-S_D^\alpha$, $S^\alpha_A-S^\alpha_B+S^\alpha_C-S_D^\alpha$, or $S^\alpha_A-S^\alpha_B-S^\alpha_C+S_D^\alpha$. One can maximize the magnitude of two out of the three sums with an $\vec S_\text{avg}=0$ configuration, whereas all magnitudes are minimized (all to zero) for a configuration with saturated $\vec S_\text{avg}$. The average magnitude of these terms decreases as $\vec S_\text{avg}$ increases. Because the magnitudes of these pocket-mixing terms determine the size of band repulsion at the avoided crossing, they also correlate with the size of the overall gap in the insulator regime. In total then, we roughly expect the gap size to shrink as $\vec S_\text{avg}$ increases. To help demonstrate this relation, Fig.~\ref{fig: gap diagram} shows the band gap size in the tetrahedral and two canted tetrahedral configurations as $J_{vc}$ and $J_{cc}$ are varied.

\section{Conclusion}

In this work, we studied an effective itinerant-electron-Kondo model on the triangular lattice with a 4-site magnetic unit cell. We calculated several magnetic phase diagrams of the model, which include chiral, in particular tetrahedral, spin ground states. One finds that the existence of chiral ground states also persists in the presence of an external magnetic field.  When in a chiral spin state, the system can exhibit a gap and form a Chern insulator with anomalous Hall conductivity $\sigma_{xy}=4\,e^2/h$. The Berry curvature in these insulators is localized near the band  avoided crossings of the lowest two sets of bands, which occur in a ring around $\bk=0$.

These results build on findings in models which are similar but which use a tight-binding description for the itinerant electrons. Our findings show that this specific description of the electron dynamics is unnecessary for realizing chiral spin order and a QAHE on the triangular lattice, and that such phenomena are present in a broader class of electron dispersions. Namely, only low-energy valence and conduction bands localized around the $\Gamma$ and $M$ points are needed. Also notable is the larger anomalous Hall conductivity  in our model compared to the tight-binding models, which have $\sigma_{xy}=e^2/h$.

This model is most relevant to 2D and quasi-2D triangular lattice materials with classically-natured magnetic moments and semiconductor or semi-metal behavior, like GdGaI. We also note that the model assumes the system is magnetically isotropic and maintains the symmetries of the triangular lattice; hence such properties are likely present in the best candidate materials.

There are several ways this work could be expanded. One interesting path would be a more expansive mapping of the spin ground state phase diagram. For instance, in our analysis, we focused on varying the inter-pocket couplings $J_{vc}$ and $J_{cc}$, while keeping the intra-pocket couplings $J_{v}$ and $J_{c}$ fixed; however, the diagrams also suggest that finite $J_{v}$ and $J_{c}$ have a role of stabilizing chiral states in certain regions. Namely, there are some points in the top left of $(J_v,J_c)=(2,4)$ diagram, Fig.~\ref{fig: Spin phase diagram 1}, which host chiral states which are otherwise coplanar in the $(J_v,J_c)=(0,0)$ diagram, Fig.~\ref{fig: Spin phase diagram 2}. Perturbatively adding spin-spin couplings, like a Heisenberg interaction, could also be a realistic consideration. Similarly, a more expansive study of the topological phase diagram could help determine the existence of other QAH states in the model. The metallic state with approximate integer AH conductivity also provides a potentially interesting scenario.

\begin{acknowledgments}

Use was made of computational facilities purchased with funds from the National Science Foundation (CNS-1725797) and administered by the Center for Scientific Computing (CSC). The CSC is supported by the California NanoSystems Institute and the Materials Research Science and Engineering Center (MRSEC; NSF DMR 2308708) at UC Santa Barbara.  This work was supported by the NSF CMMT program under Grant No. DMR-2419871, and by the Simons Collaboration on Ultra-Quantum Matter, which is a grant from the Simons Foundation (Grant No. 651440).

\end{acknowledgments}

\appendix

\section{Symmetry Constraints on $J_I^{ij}$}
\label{app: matrix constraints}
Requiring the Hamiltonian to share the spatial symmetries of the triangular lattice places constraints on the Kondo couplings $J_I^{ij}$. In this section, we see that these constraints reduce the number of independent couplings to four.

Let's first take the direct lattice basis to be $\vec a_1=(1,0),\,\vec a_2=(1/2,\sqrt{3}/{2})$. The reciprocal basis vectors are then $\vec b_1=2\pi(1,-{1}/{\sqrt3}),\,\vec b_2=2\pi(0,{2}/{\sqrt3})$, and the distinct $M$ points are $\vec M_a=({\vec b_1 + \vec b_2})/{2}$, $\vec M_b = {\vec b_1}/{2}$, and $\vec M_c={\vec b_2}/{2}$.

Under a transformation, which acts on the electrons via $\mathcal O$ and the spins as $\vec S_I\to\vec S_{I'}$, the Kondo terms transform as 
\begin{align}
\vec S_I \cdot\Phi^\dagger_{k,\alpha} J_I \vec\sigma_{\alpha\alpha'}\Phi_{k,\alpha'}\longrightarrow
\vec S_{I'} \cdot (\Phi^\dagger_{k,\alpha}\mathcal O^T) J_I \vec\sigma_{\alpha\alpha'} (\mathcal O\Phi_{k,\alpha'})\,.
\end{align}
If this transformation is a symmetry of the Hamiltonian, then the coupling matrices must satisfy
\begin{align}
\label{eq: constraint equation}
J_{I'}=\mathcal O^T J_I\mathcal O
\end{align}
for each $I=A,B,C,D$.

 The Bloch ladder operators $c_k^\dagger, c_k$ transform under $T_{\vec r}$, a translation by $\vec r$, as $c_k^\dagger\to e^{i\vec k\cdot\vec r}c_k^\dagger$ and $c_k\to e^{-i\vec k\cdot\vec r}c_k$. The spin at site $i$ transforms into the spin at site $i+\vec r$. The nearest neighbor directions are given by $\vec r=\pm\vec a_1,\, \pm\vec a_2,\, \pm(\vec a_2-\vec a_1)$; given the 4-sublattice spin pattern, we can track any translation's effect on the lattice and spins by tracking the three transformations $T_{\vec a_1}$, $T_{\vec a_2}$, and $T_{\vec a_2-\vec a_1}$. For $\vec r=\vec a_1$, we have
\begin{align}
\nonumber
&c_{a,k} \to- e^{i\vec k\cdot \vec a_1}c_{a,k}\,,\quad
c_{b,k} \to- e^{i\vec k\cdot \vec a_1}c_{b,k}\,,\\
&c_{c,k} \to e^{i\vec k\cdot \vec a_1}c_{c,k}\,,
\end{align}
where we've used that $\vec a_i\cdot \vec b_j=2\pi\delta_{ij}$. The valence state operators do not pick up another factor beyond $e^{i\vec k\cdot \vec a_1}$, since they are at the $\vec\Gamma=0$. One finds similar transformations for $\vec r=\vec a_2$ and $\vec r=\vec a_2-\vec a_1$. The matrix representations are
\begin{align}\nonumber
&\mathcal O_{T_{\vec a_1}}=
\begin{pmatrix}
    1 & & & \\
    & -1 & & \\
   & & -1 &  \\
   & & & 1
\end{pmatrix}\,,\quad
\mathcal O_{T_{\vec a_2}}=
\begin{pmatrix}
    1 & & & \\
    & -1 & & \\
   & & 1 &  \\
   & & & -1
\end{pmatrix}\,,\\
&\mathcal O_{T_{\vec a_2-\vec a_1}}=
\begin{pmatrix}
    1 & & & \\
    & 1 & & \\
   & & -1 &  \\
   & & & -1
\end{pmatrix}\,.
\end{align}
The spins transform as (in permutation notation): $(\vec S_A, \vec S_D)$, $(\vec S_B, \vec S_C)$ for $T_{\vec a_1}$, $(\vec S_A, \vec S_C)$, $(\vec S_B, \vec S_D)$ for $T_{\vec a_2}$, and $(\vec S_A, \vec S_B)$, $(\vec S_C, \vec S_D)$ for $T_{\vec a_2-\vec a_1}$.

Under an in-plane rotation $R\in SO(2)$ about the site $\vec r_0$, the Bloch ladder operator transforms as $c_k\to e^{i( R\,\vec k-\vec k)\cdot\vec r_0}\,c_{R\,\vec k}$. We find the transformation matrices and corresponding spin transformations for a $120\degree$ CCW rotation about each site $\{A,B,C,D\}$. Taking $\vec r_0=0$ to be an A site, the matrices are
\begin{align}\nonumber
&\mathcal O_{R^{120\degree}_A}=
\begin{pmatrix}
    1 & 0& 0&0 \\
    0&  0& 0& 1 \\
   0& 1 & 0&0  \\
   0& 0& 1&0 
\end{pmatrix}\,,\quad
\mathcal O_{R^{120\degree}_B}=
\begin{pmatrix}
    1 & 0& 0&0 \\
    0&  0& 0& -1 \\
   0& -1 & 0&0  \\
   0& 0& 1&0 
\end{pmatrix}\,,\\
&\mathcal O_{R^{120\degree}_C}=
\begin{pmatrix}
    1 & 0& 0&0 \\
    0&  0& 0& -1 \\
   0& 1 & 0&0  \\
   0& 0& -1&0 
\end{pmatrix}\,,\quad
\mathcal O_{R^{120\degree}_D}=
\begin{pmatrix}
    1 & 0& 0&0 \\
    0&  0& 0& 1 \\
   0& -1 & 0&0  \\
   0& 0& -1&0 
\end{pmatrix}\,,
\end{align}
and the spins transform as: $(\vec S_C,\vec S_B, \vec S_D)$ for $O_{R^{120\degree}_A}$, $(\vec S_A,\vec S_C, \vec S_D)$ for $O_{R^{120\degree}_B}$, $(\vec S_A,\vec S_D, \vec S_B)$ for $O_{R^{120\degree}_C}$, and $(\vec S_A,\vec S_B, \vec S_C)$ for $O_{R^{120\degree}_D}$.

The final lattice symmetry we consider is a mirror along the $y$-axis through an A site, whose matrix is given by
\begin{align}
\mathcal O_{M^A_{y}}=
\begin{pmatrix}
    1 & 0& 0&0 \\
    0&  0& 1& 0 \\
   0& 1 & 0&0  \\
   0& 0& 0&1 
\end{pmatrix}\,,    
\end{align}
and which has the spin transformation $(\vec S_B,\vec S_C)$.

Forming a system of constraint equations by putting these $\mathcal O$ matrices from translations, rotations, and the mirror into Eq.~\eqref{eq: constraint equation}, one finds that there are four independent couplings $\{J_v\,,J_{c}\,,J_{vc}\,,J_{cc}\}$, and that the coupling matrices are given by
\begin{align}
\label{eq: coupling matrices full}
\nonumber
&J_A=
\begin{pmatrix}
    J_{v} & J_{vc}& J_{vc}&J_{vc} \\
    J_{vc}&  J_{c}& J_{cc}& J_{cc} \\
   J_{vc}& J_{cc} & J_{c}&J_{cc}  \\
   J_{vc}& J_{cc} & J_{cc} & J_{c} 
\end{pmatrix}\,,\\\nonumber
&J_B=
\begin{pmatrix}
    J_{v} & J_{vc}& -J_{vc}&-J_{vc} \\
    J_{vc}&  J_{c}& -J_{cc}&- J_{cc} \\
   -J_{vc}& -J_{cc} & J_{c}&J_{cc}  \\
   -J_{vc}&- J_{cc} & J_{cc} & J_{c} 
\end{pmatrix}\,,\\\nonumber
&J_C=
\begin{pmatrix}
    J_{v} & -J_{vc}& J_{vc}&-J_{vc} \\
    -J_{vc}&  J_{c}& -J_{cc}& J_{cc} \\
   J_{vc}& -J_{cc} & J_{c}&-J_{cc}  \\
   -J_{vc}& J_{cc} & -J_{cc} & J_{c} 
\end{pmatrix}\,,\\
&J_D=
\begin{pmatrix}
    J_{v} & -J_{vc}& -J_{vc}&J_{vc} \\
   - J_{vc}&  J_{c}& J_{cc}& -J_{cc} \\
   -J_{vc}& J_{cc} & J_{c}&-J_{cc}  \\
   J_{vc}& -J_{cc} &- J_{cc} & J_{c} 
\end{pmatrix}\,,
\end{align}
with the column indices corresponding to the pocket indices in the order $\{v,a,b,c\}$. We observe that the remaining triangular lattice symmetries do not further constrain the couplings beyond Eq.~\eqref{eq: coupling matrices full}.

\section{Parametrization for Zero-magnetization Spin States}
\label{app: spin parametrization}

Any zero-magnetization spin configuration in the model can be parametrized as
\begin{align}
    \label{eq: spin parametrization}
    &\vec S_A=(S^x,S^y,S^z),\quad\vec S_B=(-S^x,-S^y,S^z), \\\nonumber
    &\vec S_C=(-S^x,S^y,-S^z),\quad\vec S_D=(S^x,-S^y,-S^z).
\end{align}
We assume unit normalization for each $\vec S_I$. To start, we consider the three vector combinations $\{\vec S_{AB},\vec S_{AC},\vec S_{AD}\}$, where $\vec S_{AI}\equiv \vec S_A+\vec S_I$. We will assume these are each non-zero for now. These vectors form an orthogonal set. For example, we have
\begin{align}
    \vec S_{AB}\cdot\vec S_{AC}= 1+\vec S_A\cdot\vec S_B + \vec S_A\cdot\vec S_C + \vec S_B\cdot\vec S_C\,,
\end{align}
which we will simplify with the following identity that uses zero-magnetization and unit normalization:
\begin{align}
1=|\vec S_D|^2&=|\vec S_A + \vec S_B + \vec S_C|^2 \\\nonumber
&= 3+2\left(\vec S_A\cdot\vec S_B + \vec S_A\cdot\vec S_C + \vec S_B\cdot\vec S_C \right)\,.
\end{align}
Substituting this into the dot product indeed yields $\vec S_{AB}\cdot\vec S_{AC}=0$. Similar calculations of the other dot products show that they are also zero. Using the freedom of global spin rotation, we can then change to coordinates such that $\{\vec S_{AB},\vec S_{AC},\vec S_{AD}\}$ lie along the $x,y,z$-axes. Aligning $\vec S_{AB}$ along the, say, $z$-direction and defining $\vec S_A\equiv(S^x,S^y,S^z)$ then requires $S_B^x=-S^x$ and $S_B^y=-S^y$, while $S_B^z=S^z$ is determined by normalization. The other spins are obtained similarly.

The choice $S_B^z=-S^z$ to satisfy normalization is also valid but collapses the spin state to the specific configuration $\vec S_A=-\vec S_B$ and $\vec S_C=-\vec S_D$. Such antiparallel pair states can still be described by Eq.~\eqref{eq: spin parametrization} but obtained by a different procedure. First, perform a global spin rotation so that the two pairs lie in one of the coordinate planes, e.g., to achieve $\vec S_A=-\vec S_B$ one would choose the $xy$ plane; then, use the remaining global spin rotation freedom to rotate $\vec S_A$ in the plane until the parametrization of $\vec S_C$ and $\vec S_D$ is accurate.
%\kk{ I was thinking that if $\vec{\mathbf{S}}_{AB}=0$, then $\vec{\mathbf{S}}_B=-\vec{\mathbf{S}}_A$ already follows immediately, so this may not be something achieved by a global rotation. Then the zero-magnetization condition implies $\vec{\mathbf{S}}_D=-\vec{\mathbf{S}}_C$, which should just correspond to the special case $S^z=0$ of Eq. (B1).} \kai{Sorry I'm not sure if understand your comment -- I think what's written there doesn't contradict what you're saying? If you're wondering why I wrote this extra paragraph here, it's because it's not true that any configuration $\vec{\mathbf{S}}_A=-\vec{\mathbf{S}}_B$ and $\vec{\mathbf{S}}_D=-\vec{\mathbf{S}}_C$ is automatically in the form of Eq.~\eqref{eq: spin parametrization}, and like you said, if $\vec S_{AB}=0$ then you can't align it to an axis. An example of a configuration that doesn't fall into the parametrization right away is $\vec S_A = - \vec S_B=\hat z$ and $\vec S_D=-\vec S_C=\hat x$. To get that state into the form of Eq.~\eqref{eq: spin parametrization}, you have to rotate the coordinates so that $S^x=S^y=1/\sqrt{2}$ and indeed $S^z=0$. To get the spins to satisfy Eq.~\eqref{eq: spin parametrization} I'm pretty sure you generically have to use a spin rotation like this to get into the right coordinates. }

\section{Determining the Spin Ground State}
\label{app: external field}

\begin{figure*}[!ht]
    \centering
    \includegraphics[width=1\linewidth]{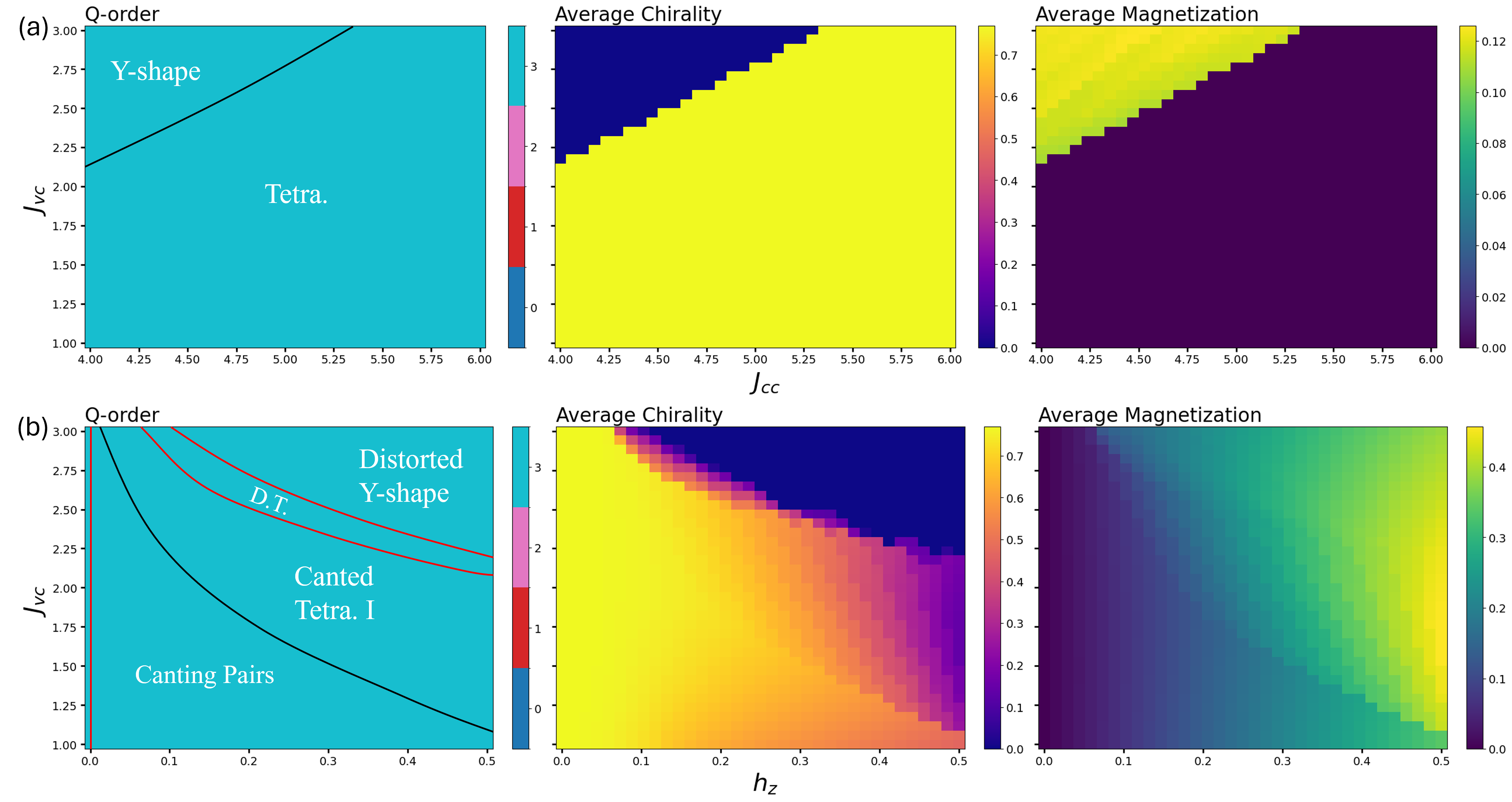}
    \caption{The $q$-ordering, average scalar chirality, and average magnetization of the ground states as $J_{vc}$ and, (a) $J_{cc}$ or (b) an external field $\vec h=h_z\,\hat z$, are varied. The other couplings are held fixed to $(J_v, J_c)=(0,0)$ in both cases and $J_{cc}=6$ in (b). Black and red lines denote first and second order phase transitions, respectively. Tetrahedral states exist at $h_z=0$ and transition to Canting Pairs at finite field. The orientation of the Distorted Tetrahedral state is the time-reversed version of the orientation in Fig.~\hyperref[fig: Spin phase diagram 1]{\ref*{fig: Spin phase diagram 1}i}. The Distorted Y-shape state is obtained by rotating (along the dashed circle) the two equally-canted legs of the D.T. state into the plane of the other two spins, hence forming a coplanar state like the Y-shape but with differing separation angles. }
    \label{fig: Spin phase diagram 2}
\end{figure*}

To account for an external magnetic field, we add to the Hamiltonian a Zeeman term coupling the field and spins:
\begin{align}
H_h\equiv -N_m \sum_{I=A}^D \vec{h}\cdot \vec S_I\,,
\end{align}
where $N_m$ is the number of 4-site magnetic unit cells in the lattice. The total energy of the system is then 
\begin{align}
&E_\text{tot}[\{\vec S_I\}] =\\\nonumber 
&\frac{A}{(2\pi)^2}\int_{|\bk|\leq\Lambda} d^2k \sum_{n}n_F(\varepsilon_n(\bk))\,\varepsilon_n(\bk) -N_m \sum_{I=A}^D \vec{h}\cdot \vec S_I\,,
\end{align}
with $A$ the total lattice area and $\varepsilon_n(\bk)$ the $n^\text{th}$ band. One can forgo using $N_m$ by noting that $A=N_m\,A_\text{muc}$, where $A_\text{muc}$ is the area of the magnetic unit cell, and writing the energy per area
\begin{align*}
&\epsilon_\text{tot}[\{\vec S_I\}]= \\\nonumber
& {\frac{1}{(2\pi)^2}}\int_{|\bk|\leq\Lambda} d^2k \sum_{n}n_F(\varepsilon_n(\bk))\,\varepsilon_n(\bk) - \frac{1}{A_\text{muc}}\sum_{I=A}^D \vec{h}\cdot \vec S_I\,.
\end{align*}
We can then use this energy density to find the spin ground state of the system.

Fig.~\ref{fig: Spin phase diagram 2} shows spin phase diagrams similar to those in the main text but in the limit where the intra-pocket Kondo couplings are zero, $(J_v, J_c)=(0,0)$. In the zero-field diagram, tetrahedral states are joined by Y-shape, or 120\degree, states, a coplanar arrangement where there are two degenerate spins and two non-degenerate spins, all separated by $120\degree$ from each other. With the external field, similar phases appear as in the $(J_v, J_c)=(2,4)$ diagrams.

\section{$\sigma_{xy}=0$ for Isotropic Conduction Bands}
\label{app: isotropic}

When the conduction bands are isotropic, $\alpha=1$, the Bloch Hamiltonian gains an extra symmetry which results in a vanishing Hall conductivity. Consider two points $\mathbf k\equiv (k_1,k_2)$ and $\tilde{\mathbf{k}}\equiv (k_2,k_1)$, related by a mirror along $k_x=k_y$. This mirror in momentum space is a symmetry of the isotropic Bloch Hamiltonian, so there exists a $\mathbf{k}$-independent unitary matrix $U_M$ such that
\begin{align}
\mathcal{H}(\tilde{\mathbf k}) = U_M\,\mathcal{H}(\mathbf k)\,U_M^\dagger\,.
\end{align}
Hence the eigenvalues satisfy $\varepsilon_n(\mathbf k)=\varepsilon_n(\tilde{\mathbf k})$, while the eigenvectors are related by
\begin{align}
|\tilde n\rangle \equiv |n(\tilde{\mathbf k})\rangle = U_M |n(\mathbf k)\rangle\,.
\end{align}
Differentiating the symmetry relation gives
\begin{align}
\partial_{k_x}\mathcal{H}(\tilde{\mathbf k}) &= U_M\,\partial_{k_y}\mathcal{H}(\mathbf k)\,U_M^\dagger\,,\\
\partial_{k_y}\mathcal{H}(\tilde{\mathbf k}) &= U_M\,\partial_{k_x}\mathcal{H}(\mathbf k)\,U_M^\dagger\,.
\end{align}
As a consequence, the Berry curvatures at the points $\mathbf k$ and $\tilde{\mathbf k}$ are related:

\begin{widetext}
\begin{align*}
\Omega_n(\tilde{\mathbf k})
&= i\sum_{m \neq n} \frac{\langle \tilde n | \partial_{k_x}\mathcal{H}(\tilde{\mathbf k}) | \tilde m \rangle \langle \tilde m | \partial_{k_y}\mathcal{H}(\tilde{\mathbf k}) | \tilde n \rangle - \langle \tilde n | \partial_{k_y}\mathcal{H}(\tilde{\mathbf k}) | \tilde m \rangle \langle \tilde m | \partial_{k_x}\mathcal{H}(\tilde{\mathbf k}) | \tilde n \rangle}{(\varepsilon_m(\tilde{\mathbf k}) - \varepsilon_n(\tilde{\mathbf k}))^2}\,, \\
&= i\sum_{m \neq n} \frac{\langle n | U_M^\dagger \partial_{k_x}\mathcal{H}(\tilde{\mathbf k}) U_M | m \rangle \langle m | U_M^\dagger \partial_{k_y}\mathcal{H}(\tilde{\mathbf k}) U_M | n \rangle - \langle n | U_M^\dagger \partial_{k_y}\mathcal{H}(\tilde{\mathbf k}) U_M | m \rangle \langle m | U_M^\dagger \partial_{k_x}\mathcal{H}(\tilde{\mathbf k}) U_M | n \rangle}{(\varepsilon_m(\mathbf k) - \varepsilon_n(\mathbf k))^2} \\
&= i\sum_{m \neq n} \frac{\langle n | \partial_{k_y}\mathcal{H}(\mathbf k) | m \rangle \langle m | \partial_{k_x}\mathcal{H}(\mathbf k) | n \rangle - \langle n | \partial_{k_x}\mathcal{H}(\mathbf k) | m \rangle \langle m | \partial_{k_y}\mathcal{H}(\mathbf k) | n \rangle}{(\varepsilon_m(\mathbf k) - \varepsilon_n(\mathbf k))^2}\,, \\
&= -\Omega_n(\mathbf k)\,.
\end{align*}
\end{widetext}

Because $\varepsilon_n(\mathbf k)=\varepsilon_n(\tilde{\mathbf k})$, both points have the same occupation, and the Berry curvatures cancel pairwise upon the momentum summation in Eq.~\eqref{eq: Hall conductivity}. Note that this is independent of the spin configuration.

\section{Explanation of $\sigma_{xy}=4\,e^2/h$}
\label{app: Chern number}

\begin{figure*}[!ht]
    \centering
    \includegraphics[width=\linewidth]{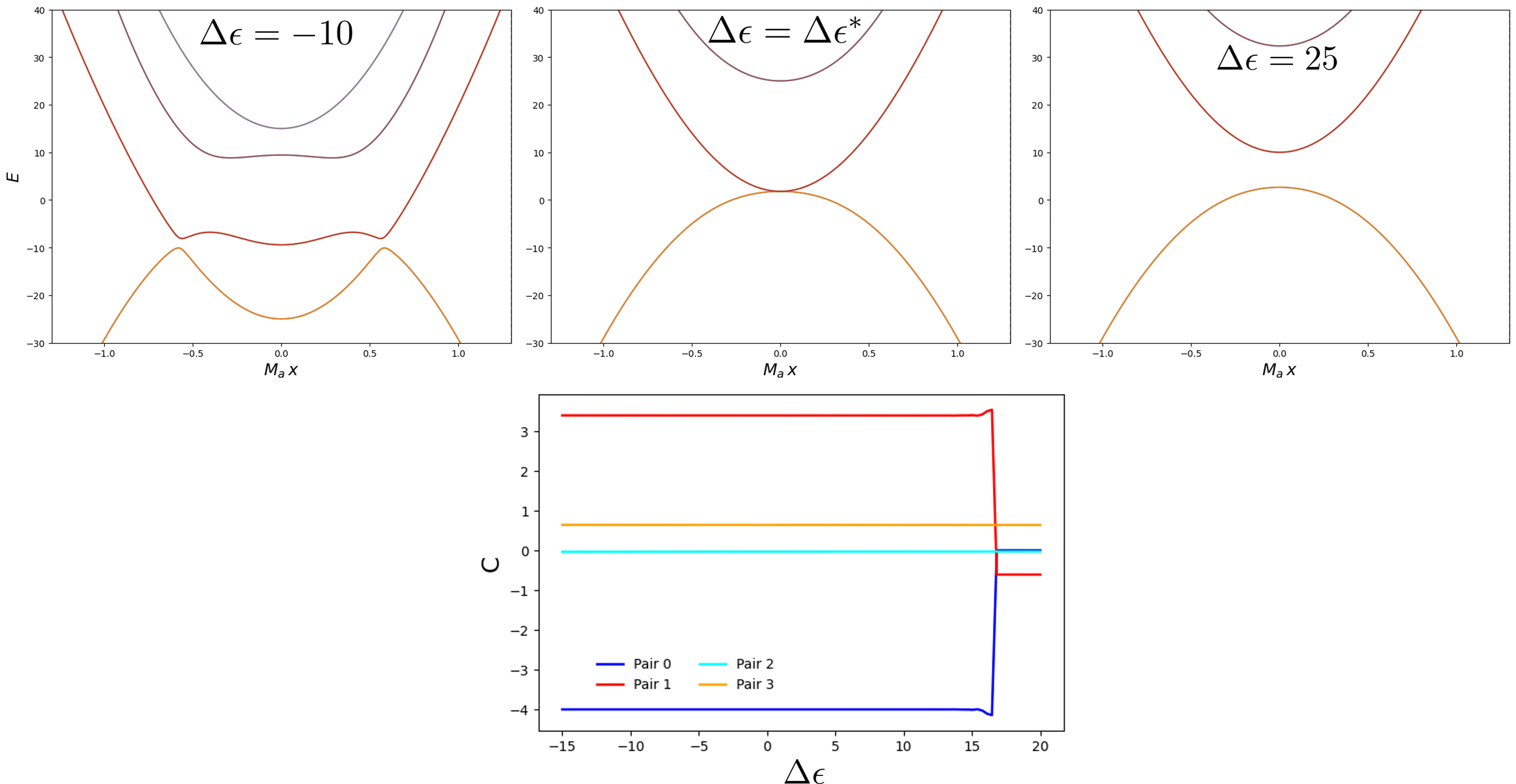}
    \caption{The top figures show several $(J_{vc},J_{cc})=(2,5)$ tetrahedral band structures as the gap between the itinerant conduction and valence bands, $\Delta\epsilon\equiv \epsilon_c-\epsilon_v$, is tuned. The lowest pair of bands in the left and right structures have Chern number $C=4$ and $C=0$, respectively. The middle band structure is at the critical point between the topological phases, which occurs at $\Delta \epsilon^\star\approx16.8$. The bottom panel shows the summed Chern number for each pair of degenerate bands as $\Delta\epsilon$ is tuned.}
    \label{fig: Chern number vs ec}
\end{figure*}

To better understand the Chern number of $|C|=4$ in the model, we study the tetrahedral bands close to the topological phase transition. For the tetrahedral configuration, the Bloch Hamiltonian block-diagonalizes into two identical non-interacting sectors (see Eq.~\eqref{eq: Hamiltonian blocks}); we study one copy at a time and double the results to describe the full system.

One can tune the topological transition with $\epsilon_c-\epsilon_v$, the gap between the itinerant conduction and valence bands. In the non-trivial regime, the Chern number originates from a band inversion occurring between the lowest two bands, whereas the upper two bands do not contribute. Fig.~\ref{fig: Chern number vs ec} shows the bands at the critical point when they touch at $\bk=0$ and on either end of the transition, as well as the band Chern numbers as functions of $\epsilon_c-\epsilon_v$.

Because the topology of the lowest band is controlled primarily by the band above it, we can analytically study the lowest band's Chern number using an effective 2-band Hamiltonian. We form this effective Hamiltonian by projecting the Bloch Hamiltonian $\hat {\mathcal H}_0$ (Eq.~\eqref{eq: H up}) onto the lowest two bands at the critical point and perturbatively expanding around $\bk=0$. We start by performing a Schrieffer-Wolff transformation and keeping to lowest order in $\bk$:
\begin{align}
    H^{4\times 4}_\text{eff} = P \hat {\mathcal H}_0P + \mathcal{O}\left(k_x,k_y\right)^4\,.
\end{align}
Here, $P=U\,U^\dagger$ is the projection onto the subspace spanned by the lowest two eigenstates $\{\ket{u_1},\ket{u_2}\}$ at $\bk=0$ at the critical point, and $U\equiv(\ket{u_1}\,\,\ket{u_2})$ is a $4\times2$ matrix. The second term in the Schrieffer-Wolff transformation is omitted since it is quartic in $k_x,k_y$. We can thus obtain the 2-band effective Hamiltonian by expressing $H^{4\times 4}_\text{eff}$ in the $\{\ket{u_1},\ket{u_2}\}$ basis:

\begin{align}
    H_\text{eff}^{2\times2}=U^\dagger \hat {\mathcal H}_0U\equiv b(\bk)\,\sigma_0+\vec d(\bk)\cdot\vec\sigma
\end{align}
with
\begin{align}
    & b(\bk)=b_0+b_1\,(k_x^2+k_y^2)\,,\\\nonumber
    & \vec d(\bk) = \Big(\lambda_0(-k_x^2+k_y^2),\, -2 \lambda_0\,k_x k_y,\, \Delta - \lambda_1 (k_x^2+k_y^2)\Big)\,.
\end{align}
Here, $b_0$, $b_1$, $\lambda_0$, and $\lambda_1$ are positive constants determined from the critical point; we add the effective mass term $\Delta$, proportional to $\epsilon_v-\epsilon_c$, to tune away from the critical point. 

The $|\bk|\leq\Lambda$ contribution to the lowest band's Chern number can be calculated via
\begin{align}
    \label{eq: Chern number}
    C = \frac{1}{4\pi}\int_{|\bk|\leq\Lambda} d^2k\,\hat d\cdot\left(\partial_{k_x}\hat d \times\partial_{k_y}\hat d \right)\,,
\end{align}
which measures how many times the pseudospin vector $\hat d(\bk)\equiv \vec d(\bk)/|\vec d(\bk)|$ wraps the Bloch sphere over the $\bk$-integral. In our case, the azimuthal winding of the pseudovector is given by 
\begin{align}
\arg(d_x+id_y)=\arg(-\lambda_0\bk^2\,e^{i2\phi})=2\phi+\pi\mod 2\pi\,,
\end{align}
meaning it wraps twice azimuthally. Additionally, for $\Delta>0$ and sufficiently large $\Lambda$, $d_z$ changes signs once going from $\bk=0$ to $|\bk|=\Lambda$. Hence, the pseudovector covers the Bloch sphere twice over the $\bk$-integral, indicating a Chern number change of $|C|=2$ at the topological transition. Since the trivial phase has $C=0$, the nontrivial phase has, when considering both sectors together, $|C|=4$.

The canted tetrahedral states are adiabatically connected to the tetrahedral one, so they maintain the same quantization as long as the gap does not close.

\bibliography{refs}

\end{document}